\documentclass[aps,preprint,nofootinbib,eqsecnum,superscriptaddress]{revtex4}
\usepackage{amsmath,amssymb,bm}
\usepackage{graphicx}
\usepackage{latexsym}

\def\be{\begin{equation}}
\def\ee{\end{equation}}
\def\bea{\begin{eqnarray}}
\def\eea{\end{eqnarray}}
\def\ba{\begin{array}}
\def\ea{\end{array}}

\newcommand{\vk}{{\bf {k}}}

\newcommand{\vko}{{\bf k_1}}

\begin{document}

\title{Quantum critical transport in clean graphene}
\author{Lars Fritz}
\affiliation{Department of Physics, Harvard University, Cambridge MA 02138, USA}
\author{J\"org Schmalian}
\affiliation{Ames Laboratory and Department of Physics and Astronomy, Iowa State
University, Ames, IA 50011, USA}
\author{Markus M\"uller}
\affiliation{Department of Physics, Harvard University, Cambridge MA 02138, USA}
\author{Subir Sachdev}
\affiliation{Department of Physics, Harvard University, Cambridge MA 02138, USA}
\date{\today\\
\vspace{0.6in}}

\begin{abstract}
We describe electrical transport in ideal single-layer graphene at zero
applied bias. There is a crossover from collisionless transport at
frequencies larger than $k_B T/\hbar$ ($T$ is the temperature) to
collision-dominated transport at lower frequencies. The d.c. conductivity is
computed by the solution of a quantum Boltzmann equation. 
Due to a logarithmic singularity in the collinear scattering amplitude (a consequence of 
relativistic dispersion in two dimensions) 
quasi-particles and -holes moving in the same direction tend to an effective 
equilibrium distribution whose parameters depend on the direction of motion.
This property allows us to find the non-equilibrium distribution functions and 
the quantum critical conductivity exactly to leading order in $1/|\log(\alpha)|$ 
where $\alpha$ is the coupling constant characterizing the Coulomb interactions.
\end{abstract}

\maketitle

\section{Introduction}

Despite the intense experimental and theoretical interest in the electronic
properties of graphene \cite{rmp}, there has been relatively little progress
in measuring and understanding the role of electron-electron interactions. 
However, the recent ability to grow ultrahigh mobility, suspended, single
layer graphene \cite{stormer,andrei} promises that the situation may well
change in the near future.

This paper will examine the role of electron-electron interactions in an
infinite sample of single layer graphene without impurities. We will also
restrict our attention to the undoped case, so that the chemical potential
is at the node of the massless Dirac spectrum. Our results can be extended
to include a non-zero chemical potential and a dilute concentration of
impurities: this was discussed recently in Ref.~\onlinecite{mhd} for a
low-frequency `hydrodynamic' regime, and additional results will appear in
forthcoming work.

The key to understanding electron-electron interactions in clean, undoped
graphene is the fact that it is a nearly `quantum critical' system with
marginally irrelevant Coulomb interactions \cite{son,guinea,ys, gorbar,joerg}%
. This implies that the inelastic, electron-electron scattering rate is of
order $k_{B}T/\hbar $, where $T$ is the absolute temperature, and there is a
crossover from `hydrodynamic' to `collisionless' transport as the
measurement frequency ($\omega $) is increased past the scattering rate \cite%
{damless,ssqhe}. These two regimes are captured in the following limiting
forms for the frequency dependence of the electrical conductivity, $\sigma $%
,
\begin{equation}
\sigma (\omega )=\left\{
\begin{array}{ccc}
\displaystyle\frac{e^{2}}{h}\left[ \frac{\pi }{2}+\mathcal{O}\left( \frac{1}{%
\ln (\Lambda /\hbar \omega )}\right) \right] & ~,~ & \hbar \omega \gg k_{B}T
\\
&  &  \\
\displaystyle\frac{e^{2}}{h\alpha ^{2}(T)}\left[ 0.760+\mathcal{O}\left(
\frac{1}{|\ln (\alpha(T))|}\right) \right] & ~,~ & \displaystyle%
\hbar \omega \ll k_{B}T \alpha^2 (T)%
\end{array}%
\right. ,  \label{res}
\end{equation}%
where $\alpha (T)$ is a temperature-dependent, dimensionless
`fine-structure constant', which controls the strength of the
electron-electron interactions (defined more precisely in Section~\ref%
{sec:rg}), and $\Lambda $ is a cutoff energy scale of the order of the 
electronic
bandwidth. The high frequency result above (the `collisionless' regime) was
obtained in Refs.~\onlinecite{joerg,herbut}. The leading term is the
conductivity of $4$ species of free massless Dirac fermions. Herbut \emph{et
al.} \cite{herbut} also obtained the coefficient of the subleading $[\ln
(\Lambda /\hbar \omega )]^{-1}$ term. The low frequency result in Eq.~(\ref%
{res}), which is the collision-dominated hydrodynamic regime, is the primary
new result of this paper. At asymptotically low temperatures we have (see
Eq.~(\ref{log}))
\begin{equation}
\alpha (T)\approx \frac{4}{\ln (\Lambda /T)};
\end{equation}%
the resulting logarithmic increase of $\sigma $ with decreasing $T$ is similar 
to those of
quantum critical systems in their upper-critical dimension \cite{ssbook},
and the inelastic scattering rate of the carriers is of order $(k_{B}T/\hbar
)\alpha ^{2}(T)$.

Related results have been obtained recently by Kashuba \cite{kashuba} in a
preprint which appeared while our paper was being completed.

We also note that our results are obtained in the context of a solution of the 
quantum Boltzmann equation. Going beyond the Boltzmann approximation, 
and in a system with perfect momentum conservation, we have to consider potentially
singular hydrodynamic ``long-time tails'' in a mode-coupling theory \cite{pavel}, 
which could modify the low frequency
behavior of the conductivity. Such effects are however innocuous here, because the long-range
Coulomb interaction suppresses density fluctuations.

\section{Renormalization group analysis}

\label{sec:rg}

Here, and in the remainder of the paper, we set $\hbar=k_B = 1$.

We begin introducing the low-energy theory for graphene, and reviewing its
renormalization group (RG) properties. The theory is expressed in terms of $%
N=4$ species of two-component Dirac fermions $\Psi _{a}$ ($a=1\ldots N$) and
the Euclidean partition function
\begin{eqnarray}
\mathcal{Z} &=&\int \mathcal{D}\Psi _{\alpha }\,\mathcal{D}A_{\tau }\,\exp
\left( -\mathcal{S}\right) ,  \notag \\
\mathcal{S} &=&\sum_{a=1}^{N}\int d\mathbf{x}\int d\tau \,\Psi _{a}^{\dagger
}(\mathbf{x},\tau )\Biggl[\frac{\partial }{\partial \tau }+ieA_{\tau }(%
\mathbf{x},\tau )+iv_{F}^{0}\sigma ^{x}\left( \frac{\partial }{\partial x}+i%
\frac{e}{c}A_{x}\right)  \notag \\
&~&~~~~~~~~~~+iv_{F}^{0}\sigma ^{y}\left( \frac{\partial }{\partial y}+i%
\frac{e}{c}A_{y}\right) \Biggr]\Psi _{a}(\mathbf{x},\tau )+\frac{1}{2}\int
\frac{d^{2}q}{4\pi ^{2}}\int d\tau \,\frac{\varepsilon q}{2\pi }\left\vert
A_{\tau }(\mathbf{q},\tau )\right\vert ^{2}.  \notag  \label{zz}
\end{eqnarray}%
The functional integral is over fields defined in two spatial dimensions $%
\mathbf{x}=(x,y)$ and imaginary time $\tau $, $\sigma ^{x,y}$ are Pauli
matrices acting on the sublattice space of the honeycomb lattice, and 
$v_{F}^{0}$ is the bare Fermi
velocity. The scalar potential, $A_{\tau }$, mediates the $%
e^{2}/(\varepsilon |\mathbf{x}|)$ Coulomb interaction between the electrons,
where $\varepsilon =\left( \varepsilon _{A}+\varepsilon _{B}\right) /2$ is
the dielectric constant for a \ graphene sheet confined between two
dielectrica with dielectric constants $\varepsilon _{A}$ and $\varepsilon
_{B}$, respectively. We have also introduced an non-fluctuating external
vector potential $\mathbf{A}=(A_{x},A_{y})$ as a source field: this allows us
to extract the electrical current.

The renormalization group properties of $\mathcal{Z}$ have been discussed
elsewhere \cite{son,guinea,ys, gorbar,joerg}. The fermion field $\Psi _{a}$
undergoes a wavefunction renormalization, the charge $e$ remains
unrenormalized, and the velocity $v_{F}$ renormalizes to larger values with
decreasing energy scale. For the velocity renormalization, we have the RG
equation
\begin{equation}
\frac{dv_{F}}{d\ell }=f(\alpha )v_{F},
\end{equation}%
where the running \textquotedblleft fine-structure constant" is
\begin{equation}
\alpha \equiv \frac{e^{2}}{\varepsilon v_{F}}, \label{fine}
\end{equation}%
and the function $f(\alpha )=\alpha /4$ in the perturbative regime of small $%
\alpha $. We can re-express these results in terms of the RG equation for
the dimensionless coupling $\alpha $
\begin{equation}
\frac{d\alpha }{d\ell }=-\frac{\alpha ^{2}}{4}+\mathcal{O}(\alpha ^{3}).
\label{rg}
\end{equation}%
Notice that $\alpha $ scales to small values at small energies, and this is
what facilitates the transport analysis of this paper. It has been shown
that $\alpha =0$ is the only fixed point in an analysis which, in the large $N$ 
limit, also remains valid for large
values of $\alpha $~\cite{ys,son}.

We are only interested here in observables related to the electrical
current, and so we will not need the explicit form of the wavefunction
renormalization. The current is obtained by taking a functional derivative
with respect to $\mathbf{A}$, and this is protected by gauge invariance to
have the same form when expressed in terms of either the bare or
renormalized quantities~\cite{ssbook}, which we will use explicitly in 
(\ref{defj1},\ref{defj2}) below. For two dimensional graphene this
implies that the scaling dimension\ of the conductivity is exactly zero and
is unaffected by wavefunction renormalizations. This result can also
be obtained explicitly by exploring charge conservation of the system along
with the related Ward identity~\cite{joerg} and holds to arbitrary order in
perturbation theory.

We are interested here in the collision-dominated transport regime, where
the characteristic energy of excitations is $k_{B}T$. We thus use the RG
equation to scale down from some high energy cutoff scale, $\Lambda$, to a scale 
$%
k_{B}T $. Integrating Eq.~(\ref{rg}) over this regime, we obtain
\begin{equation}
\alpha (T)=\frac{\alpha ^{0}}{1+(\alpha ^{0}/4)\ln (\Lambda /T)}\stackrel{T\to 
0}{\sim} \frac{4}{\ln (\Lambda /T)}\,,
\label{log}
\end{equation}%
where $\alpha ^{0}$ is the bare value dependent upon $v_{F}^{0}$. Son \cite%
{son} has also examined the structure of the RG flow at strong coupling in
the large $N$ limit; he finds that there is a significant intermediate
energy scale over which
\begin{equation}
\alpha (T)\sim \left( \frac{T}{\Lambda }\right) ^{4/(\pi ^{2}N)}.
\label{power}
\end{equation}%
Both Eqs.~(\ref{log}) and (\ref{power}) predict a slow flow with decreasing 
temperature 
towards weak coupling. We can also use $\alpha (T)$ to obtain a $T$-dependent 
velocity
\begin{equation}
v_{F}(T)=v_{F}^{0}\frac{\alpha ^{0}}{\alpha (T)} = v_{F}^{0}\left[ 
1+\frac{\alpha^0}{4}\ln(\Lambda/T)\right].
\end{equation}

We also note that the leading order flow in $\alpha$ in Eq.~(\ref{rg})
represents an exchange-correlation effect. Ordinary screening effects are formally
higher order, and can be accounted for in the random-phase approximation by the mapping \cite{valeri,rudro}
\begin{equation}
\alpha (T) \rightarrow \frac{\alpha (T)}{1 + N \pi \alpha(T)/8}.
\end{equation}

\section{Collision-dominated transport}

\label{sec:qbe}

After initially renormalizing down to a scale $T$ as described in Eq.~(\ref%
{sec:rg}), we can now investigate the transport quantities in the
renormalized theory. So all subsequent appearances of the field $\Psi$, the
velocity $v_F$, and the coupling $\alpha$ implicitly refer to the $T$%
-dependent renormalized quantities obtained as described in Section~\ref%
{sec:rg}. We will not explicitly write-out this $T$ dependence.

Our formulation of the transport properties of the renormalized theory of
weakly-interacting massless Dirac fermions closely follows that presented in
Ref.~\onlinecite{ssqhe}. This previous work considered massless Dirac
fermions interacting with a weak statistical interaction due to a
Chern-Simons term, and here we only need to replace the Chern-Simons term by
a Coulomb interaction. The transport analysis is easiest in the real-time
operator formulation with the Hamiltonian
\begin{eqnarray}
H &=& H_0 + H_1  \notag \\
H_0 &=& \int d \mathbf{x} \left[ v_F \Psi_a^{\dagger} \left( -i\sigma^i
\partial_i \right) \Psi_a \right] \\
H_1 &=& \frac{1}{2} \int \frac{d^2 k_1 }{(2 \pi )^2} \frac{d^2 k_2 }{(2 \pi
)^2} \frac{d^2 q }{(2 \pi )^2}\Psi_a^{\dagger} ( \mathbf{k}_2-\mathbf{q} )
\Psi_a ( \mathbf{k}_2 ) V({\bf q}) \Psi_b^{\dagger} (
\mathbf{k}_1+\mathbf{q} ) \Psi_b ( \mathbf{k}_1 )\,,
\end{eqnarray}
with the Coulomb interaction
\begin{eqnarray}
\label{Cb}
V({\bf q})= \frac{2 \pi e^2}{\varepsilon |{\bf q}|},
\end{eqnarray}
and $a = 1, \ldots, N$ labeling the "flavors" of fermions ($N=4$ in graphene, 
accounting for 2 valleys and 2 spin projections).
Even though we compute our results specifically for the Coulomb interactions 
(\ref{Cb}), the formalism carries through in exactly the same manner for 
arbitrary isotropic two body potentials. 

The simplest formulation of the transport equations is in a basis which
diagonalizes the Hamiltonian $H_{0}$. To do this, we first express $\Psi $
in its Fourier components
\begin{equation}
\Psi _{a}(\mathbf{x},t)=\int \frac{d^{2}k}{(2\pi )^{2}}\left(
\begin{array}{c}
c_{1a}(\mathbf{k},t) \\
c_{2a}(\mathbf{k},t)%
\end{array}%
\right) e^{i\mathbf{k}\cdot \mathbf{x}},
\end{equation}%
and then perform a unitary transformation from the Fourier mode operators $%
(c_{1a},c_{2a})$ to $(\gamma _{+a},\gamma _{-a})$:
\begin{eqnarray}
c_{1a}(k) &=&\frac{1}{\sqrt{2}}(\gamma _{+a}(\mathbf{k})+\gamma _{-a}(%
\mathbf{k}))  \notag  \label{eq:unitary} \\
c_{2a}(k) &=&\frac{K}{\sqrt{2}k}(\gamma _{+a}(\mathbf{k})-\gamma _{-a}(%
\mathbf{k})).
\end{eqnarray}%
We have introduced here a notational convention that we shall find quite
useful in the following: as $\mathbf{k}$ is a two-dimensional momentum, we
can define the complex number $K$ by
\begin{equation}
K\equiv k_{x}+ik_{y}~~~~\mbox{where}~~~~~\mathbf{k}\equiv (k_{x},k_{y})
\end{equation}%
and $k=|\mathbf{k}|=|K|$. Expressing the Hamiltonian $H_{0}$ in terms of $%
\gamma _{\pm }$, we obtain the simple result
\begin{equation}
H_{0}=\sum_{\lambda ,a}\int \frac{d^{2}k}{(2\pi )^{2}}\lambda v_{F}k\, \gamma
_{\lambda a}^{\dagger }(\mathbf{k})\gamma _{\lambda a}(\mathbf{k}),
\end{equation}%
where the sum over $\lambda $ extends over $+,-$.

Let us also express the interaction Hamiltonian $H_1$ in terms of the $%
\gamma_{\lambda a}$:
\begin{eqnarray}
&& H_1 = \sum_{\lambda_1 \lambda_2 \lambda_3 \lambda_4} \int \frac{d^2 k_1 }{%
(2 \pi )^2} \frac{d^2 k_2 }{(2 \pi )^2} \frac{d^2 q }{(2 \pi )^2}  \notag \\
&&~~~~~~~~~~~~~~~~ \times T_{\lambda_1 \lambda_2 \lambda_3 \lambda_4} (%
\mathbf{k}_1 , \mathbf{k}_2 , \mathbf{q} ) \gamma_{\lambda_4 b}^{\dagger} (
\mathbf{k}_1+\mathbf{q} ) \gamma_{\lambda_3 a}^{\dagger} ( \mathbf{k}_2-%
\mathbf{q} ) \gamma_{\lambda_2 a} ( \mathbf{k}_2 ) \gamma_{\lambda_1 b} (
\mathbf{k}_1 )
\end{eqnarray}
where
\begin{equation}
T_{\lambda_1 \lambda_2 \lambda_3 \lambda_4} (\mathbf{k}_1 , \mathbf{k}_2 ,
\mathbf{q}) = \frac{V({\bf q})}{8} \left[ 1 +
\lambda_1 \lambda_4 \frac{(K_1^{\ast} + Q^{\ast}) K_1}{|\mathbf{k}_1 +
\mathbf{q}| k_1} \right] \left[1 + \lambda_2 \lambda_3 \frac{(K_2^{\ast} -
Q^{\ast}) K_2 }{|\mathbf{k}_2 - \mathbf{q}| k_2} \right] . \label{deft}
\end{equation}

Finally, we also express the electrical current, obtained by taking a
functional derivative of the action with respect to $\mathbf{A}$, in terms
of the $\gamma _{\pm }$. For the case of a spatially independent current
(which is the only case of interest here), the result can be written as
\begin{equation}
\mathbf{J}=\mathbf{J}_{I}+\mathbf{J}_{II}
\end{equation}%
with
\begin{equation}
\mathbf{J}_{I}=ev_{F}\sum_{\lambda a}\int \frac{d^{2}k}{(2\pi )^{2}}\frac{%
\lambda \mathbf{k}}{k}\gamma _{\lambda a}^{\dagger }(\mathbf{k})\gamma
_{\lambda a}(\mathbf{k})\,,  \label{defj1}
\end{equation}%
and
\begin{equation}
\mathbf{J}_{II}=-iev_{F}\int \frac{d^{2}k}{(2\pi )^{2}}\frac{(\hat{\mathbf{z}%
}\times \mathbf{k})}{k}\left[ \gamma _{+a}^{\dagger }(\mathbf{k})\gamma
_{-a}(\mathbf{k})-\gamma _{-a}^{\dagger }(\mathbf{k})\gamma _{+a}(\mathbf{k}%
)\right]\,, \label{defj2}
\end{equation}%
where $\hat{\mathbf{z}}$, a unit vector orthogonal to the $x,y$ plane. 
$\mathbf{J}_{I}$ measures the current carried by motion of the
quasiparticles and quasiholes---notice the $\lambda $ prefactor, indicating
that these excitations have opposite charges. The operator $\mathbf{J}_{II}$
creates a quasiparticle-quasihole pair, and in the low 
frequency limit of interest here we 
may neglect $\mathbf{J}_{II}$.
Similarly to the problems
studied in Refs.[\onlinecite{damless,ssqhe}], a current carying state with
holes and electrons moving in opposite directions is consistent with a
vanishing total momentum. Thus a finite conductivity does not require the total 
momentum of the problem to relax. This is the physical reason why at the 
particle hole symmetric point, i.e., at vanishing  deviation of the chemical 
potential from the Dirac point, the d.c. conductivity is finite even in the 
absence of momentum relaxing impurities. However, as we will see below at finite 
deviation from particle hole symmetry a driving electric field always excites 
the system into a state with finite momentum. The latter cannot decay which 
entails an infinite d.c. conductivity, in accordance with the hydrodynamic 
analysis~\cite{mhd}.

We can now write down the collisionless transport equations for the
excitations. As a first step, we define the distribution functions
\begin{equation}
f_{\lambda }(\mathbf{k},t)=\left\langle \gamma _{\lambda a}^{\dagger }(%
\mathbf{k},t)\gamma _{\lambda a}(\mathbf{k},t)\right\rangle .  \label{defg}
\end{equation}%
where there is no sum over $a$ on the rhs, and we assume the distribution
functions to be the same for all valleys and spins. In equilibrium, i.e., in the
absence of external perturbations, these are related to the Fermi function
\begin{eqnarray}
f_{+}(\mathbf{k},t) &=&f^{0}(v_{F}k)= \frac{1}{e^{(v_F k -\mu)/T}+1}  
\notag \\
f_{-}(\mathbf{k},t) &=&f^{0}(-v_{F}k) = 
\frac{1}{e^{(-v_F k-\mu)/T}+1},
\end{eqnarray}%
where we temporarily allow for a finite chemical potential $\mu$.

Then to first order, in the presence of an external electric field $\mathbf{E%
}$, we find the simple equations
\begin{equation}
\left( \frac{\partial }{\partial t}+e\mathbf{E}\cdot \frac{\partial }{%
\partial \mathbf{k}}\right) f_{\lambda }(\mathbf{k},t)=0.  \label{trans0}
\end{equation}%
It is a simple matter to solve (\ref{trans0}) in linear response. First we
parameterize the change in $f_{\lambda }$ from its equilibrium value by \cite%
{yaffe}
\begin{equation}
f_{\lambda }(\mathbf{k},\omega )=2\pi \delta (\omega )f^{0}(\lambda
v_{F}k)+ e\frac{\mathbf{k}\cdot \mathbf{E}(\omega )}{k}f^{0}(\lambda
v_{F}k)(1-f^{0}(\lambda v_{F}k))g_\lambda(k,\omega ),  \label{paramet}
\end{equation}%
where we have performed a Fourier transform in time to frequencies, $\omega $%
, and introduced the unknown function $g_\lambda(k,\omega )$. 
At the particle hole symmetric point ($\mu=0$), an applied electric field 
generates an deviations in the distribution functions having opposite sign for 
quasiparticles and
quasiholes, 
\bea
\label{phsym}
g_\lambda(k,\omega)=\lambda g(k,\omega).
\eea
This reflects the fact that there is an increased number of quasiholes and 
quasiparticles  moving parallel and antiparallel to field, respectively. As 
quasiparticles and -holes have opposite charges, their electrical currents are
equal, while their net momenta have opposite signs. 

Inserting (\ref{paramet}) into (\ref{trans0}), we obtain a simple solution for 
the function $g$
\begin{equation}
g(k,\omega )=\frac{v_{F}/T}{(-i\omega +\eta )},
\end{equation}%
where $\eta $ is a positive infinitesimal. Inserting this result into (\ref%
{defg}) and (\ref{defj1}), we obtain the conductivity
\begin{eqnarray}
{\sigma }(\omega ) =\frac{\left\langle J_{I}\right\rangle}{E(\omega)} 
&=&2N\frac{e^{2}v_{F}}{(-i\omega +\eta )}\int \frac{d^{2}k%
}{(2\pi )^{2}}\frac{k_{x}^{2}}{k^{2}}\left( -\frac{\partial f^{0}(v_{F}k)}{%
\partial k}\right)  \notag \\
&=&\frac{e^{2}}{h}\frac{Nk_{B}T\ln 2}{(-i\hbar \omega +\eta )},  \label{eqp1}
\end{eqnarray}%
where, in the last equation, we have re-inserted factors of $\hbar $ and $%
k_{B}$. Note that all factors of $v_{F}$ have cancelled out: this is a
consequence of the conductivity having scaling dimension $d-2$ (where $d$ is
the spatial dimensionality), and being independent of the dynamic critical
exponent $z$. So in this free electron approximation, the real part of the
low frequency $\sigma $ is a delta function at $\omega =0$ with weight of
order $k_{B}T$.

Including interband transitions
the real part of the conductivity becomes
\begin{equation}
{\rm Re}\,\sigma \left( \omega \right) =\frac{e^{2}}{h}N\pi \left[ [ k_{B}T\ln
2] \, \delta \left( \hbar \omega \right) +\frac{1}{8}\tanh \left( \frac{\hbar
\omega }{4k_{B}T}\right) \right]  \label{resigma}
\end{equation}%
with high frequency limit ${\rm Re}\,\sigma \left( \omega \gg k_{B}T/\hbar
\right) \rightarrow e^{2}N\pi /\left( 8h\right) $. In the collisionless
regime this constant value remains the leading contribution to the
conductivity even if one includes the electron-electron Coulomb interaction%
\cite{joerg,herbut}. Next order corrections are of the form%
\begin{equation}
{\rm Re}\,\sigma \left( \omega \gg k_{B}T/\hbar \right) 
=\frac{e^{2}}{h}\frac{N\pi}{8} \left[ 1+\mathcal{O}\left( \alpha \left( \omega 
\right) \right) %
\right] ,
\end{equation}
where $\alpha (\omega )\approx 4/\ln (\Lambda /\hbar \omega )$ is the
renormalized frequency-dependent `fine-structure constant' for $k_{B}T\ll
\hbar \omega \ll \Lambda $. For $N=4$, this yields the result given in the
upper row of Eq.~(\ref{res}). Thus, in the collisionless regime, interactions
only lead to very small changes of the conductivity. In the following
subsections, we will discuss the opposite, collision dominated limit, $\hbar
\omega \ll k_{B}T$, and determine how collisions broaden the delta function
of Eq.~(\ref{resigma}) to a Drude peak.

\subsection{Quantum Boltzmann equation}

\label{cdom}

\begin{figure}[ht]
\centering
\includegraphics*[width=\textwidth]{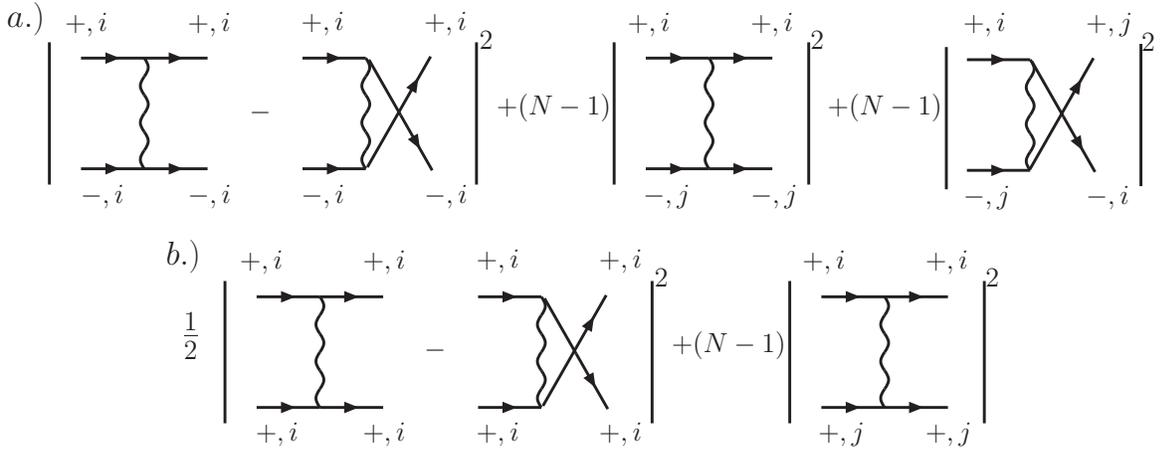}
\caption{\label{fig:goldenrule} Illustration of the Golden rule diagrams 
entering the collision term. The diagrams (a) describe scattering of oppositely 
charged particles corresponding to the term $R_1$, while the diagrams (b) 
describe scattering of like particles corresponding to the term $R_2$. Note that 
the vertex preserves the flavor $a=i,j$, but not the particle/hole nature 
$\lambda=\pm$. The factor $1/2$ of the first diagram accounts for the symmetry 
factor associated with having two indistinguishable particles in the final 
state.}
\end{figure}

We now include collision terms on the right hand side of (\ref{trans0}). We
can determine these terms by application of Fermi's golden rule~\cite{ssqhe}%
, or by the explicit derivation presented in the Appendix:
\begin{eqnarray}
&&\left( \frac{\partial }{\partial t}+e\mathbf{E}\cdot \frac{\partial }{%
\partial \mathbf{k}}\right) f_{\lambda }(\mathbf{k},t)=-\frac{(2\pi )}{v_{F}}%
\int \frac{d^{2}k_{1}}{(2\pi )^{2}}\frac{d^{2}q}{(2\pi )^{2}}\Biggl\{  \notag
\\
&&\delta (k-k_{1}-|\mathbf{k}+\mathbf{q}|+|\mathbf{k}_{1}-\mathbf{q}|)R_{1}(%
\mathbf{k},\mathbf{k}_{1},\mathbf{q})\Bigl\{f_{\lambda }(\mathbf{k}%
,t)f_{-\lambda }(\mathbf{k}_{1},t)[1-f_{\lambda }(\mathbf{k}+\mathbf{q},t)]
\notag \\
&&~~~~~~\times \lbrack 1-f_{-\lambda }(\mathbf{k}_{1}-\mathbf{q}%
,t)]-[1-f_{\lambda }(\mathbf{k},t)][1-f_{-\lambda }(\mathbf{k}%
_{1},t)]f_{\lambda }(\mathbf{k}+\mathbf{q},t)f_{-\lambda }(\mathbf{k}_{1}-%
\mathbf{q},t)\Bigr\}  \notag \\
&&\delta (k+k_{1}-|\mathbf{k}+\mathbf{q}|-|\mathbf{k}_{1}-\mathbf{q}|)R_{2}(%
\mathbf{k},\mathbf{k}_{1},\mathbf{q})\Bigl\{f_{\lambda }(\mathbf{k}%
,t)f_{\lambda }(\mathbf{k}_{1},t)[1-f_{\lambda }(\mathbf{k}+\mathbf{q},t)] \\
&&~~~~~~\times \lbrack 1-f_{\lambda }(\mathbf{k}_{1}-\mathbf{q}%
,t)]-[1-f_{\lambda }(\mathbf{k},t)][1-f_{\lambda }(\mathbf{k}%
_{1},t)]f_{\lambda }(\mathbf{k}+\mathbf{q},t)f_{\lambda }(\mathbf{k}_{1}-%
\mathbf{q},t)\Bigr\}\Biggr\}.  \notag  \label{trans1}
\end{eqnarray}%
where
\begin{eqnarray}
R_{1}(\mathbf{k},\mathbf{k}_{1},\mathbf{q}) &=&4\left( \bigl|T_{+--+}(%
\mathbf{k},\mathbf{k}_{1},\mathbf{q})-T_{+-+-}(\mathbf{k},\mathbf{k}_{1},-%
\mathbf{k}-\mathbf{q}+\mathbf{k}_{1})\bigr|^{2}\right.  \notag \\
&&\left. ~+(N-1)\bigl|T_{+--+}(\mathbf{k},\mathbf{k}_{1},\mathbf{q})\bigr|%
^{2}+(N-1)\bigl|T_{+-+-}(\mathbf{k},\mathbf{k}_{1},-\mathbf{k}-\mathbf{q}+%
\mathbf{k}_{1})\bigr|^{2}\right)\,,  \notag \\
R_{2}(\mathbf{k},\mathbf{k}_{1},\mathbf{q}) &=&4\left( \frac{1}{2}\bigl|%
T_{++++}(\mathbf{k},\mathbf{k}_{1},\mathbf{q})\ -T_{++++}(\mathbf{k},\mathbf{%
k}_{1},\mathbf{k}_{1}-\mathbf{k}-\mathbf{q})\ \bigr|^{2}\right.  \notag \\
&~&~~~~\left. ~~~~+(N-1)\bigl|T_{++++}(\mathbf{k},\mathbf{k}_{1},\mathbf{q})%
\bigr|^{2}\right)\,,  \label{deft12}
\end{eqnarray}%
which are illustrated in Fig.~\ref{fig:goldenrule}.
The terms proportional to $R_{1}$ represent collisions between oppositely
charged particles, while those proportional to $R_{2}$ are collisions
between like charges. There are also processes where a particle-hole pair is
created: as in Refs.~\onlinecite{damless,ssqhe}, these can be dropped
because they have vanishing phase space upon imposition of the energy
conservation constraint with dispersion $\varepsilon _{k}=v_{F}k$

We now proceed to the linearization of (\ref{trans1}) by inserting the
parametrization (\ref{paramet}) and find

\begin{eqnarray}
&& \frac{(-i \omega g_\lambda(k,\omega) - \lambda v_F/T)}{(e^{v_F k/T} + 1) 
(e^{-v_F k/T} +
1)} \frac{\mathbf{k}}{k}= - \frac{(2 \pi)}{v_F} \int \frac{d^2 k_1}{(2 \pi)^2%
} \frac{d^2 q}{(2 \pi)^2} \Biggl\{  \notag \\
&&~~ \frac{\delta(k -k_1 - |\mathbf{k} + \mathbf{q}| + |\mathbf{k}_1 -
\mathbf{q}|) R_1 (\mathbf{k}, \mathbf{k}_1, \mathbf{q})}{(e^{-v_F k/T} +
1)(e^{v_F k_1/T}+1) (e^{v_F |\mathbf{k} + \mathbf{q}|/T} + 1) (e^{-v_F |%
\mathbf{k}_1 - \mathbf{q}|/T} + 1)}  \notag \\
&& \times \left[ \frac{\mathbf{k}}{k} g_\lambda(k, \omega) + 
\frac{\mathbf{k}_1}{k_1}
g_{-\lambda} (k_1, \omega) - \frac{ (\mathbf{k} + \mathbf{q})}{|\mathbf{k} + 
\mathbf{q}|%
} g_\lambda (|\mathbf{k} + \mathbf{q}|, \omega) - \frac{ (\mathbf{k}_1 - 
\mathbf{q})%
}{|\mathbf{k}_1 - \mathbf{q}|} g_{-\lambda} (|\mathbf{k}_1 - \mathbf{q}|, 
\omega) %
\right]  \notag \\
&&+ \frac{\delta(k + k_1 - |\mathbf{k} + \mathbf{q}| - |\mathbf{k}_1 -
\mathbf{q}|) R_2 (\mathbf{k}, \mathbf{k}_1, \mathbf{q})}{(e^{-v_F k/T} +
1)(e^{-v_F k_1/T}+1) (e^{v_F |\mathbf{k} + \mathbf{q}|/T} + 1) (e^{v_F |%
\mathbf{k}_1 - \mathbf{q}|/T} + 1)} \label{trans2} \\
&& \times \left[ \frac{\mathbf{k}}{k} g_\lambda(k, \omega) + 
\frac{\mathbf{k}_1}{k_1}
g_\lambda(k_1, \omega) - \frac{(\mathbf{k} + \mathbf{q})}{|\mathbf{k} + 
\mathbf{q}|}
g_\lambda(|\mathbf{k} + \mathbf{q}|, \omega) - \frac{ (\mathbf{k}_1 - 
\mathbf{q})}{|%
\mathbf{k}_1 - \mathbf{q}|} g_\lambda(|\mathbf{k}_1 - \mathbf{q}|, \omega) 
\right] %
\Biggr\}\,. \nonumber 
\end{eqnarray}

The remainder of this paper is focused on the solution of the linearized
transport equation in Eq.~(\ref{trans2}) for the function $g$. It is useful
at this point to recall some crucial mathematical properties of such
transport equations, reviewed, \emph{e.g.\/}, by Ziman \cite{ziman} and Arnold
\emph{et al.} \cite{yaffe}. We can view the right hand side of Eq.~(\ref%
{trans2}) as a linear operator, the so-called collision operator $\mathcal{C}
$, acting on the function $(\mathbf{k}/k)g(k)$; we drop the implicit $\omega
$ dependence because $\mathcal{C}$ is independent of $\omega $. A key
property of $\mathcal{C}$ is that it is Hermitian with respect to the
natural inner product 
\begin{equation}
\label{innerprod}
\langle g_{1}|g_{2}\rangle \equiv \sum_\lambda \int \frac{d^{2}k}{(2\pi )^{2}}%
g_{1,\lambda}(k)g_{2,\lambda}(k).
\end{equation}%
This Hermiticity follows \cite{yaffe} from symmetry properties of $R_{1}$
and $R_{2}$ under exchanges between incoming and outgoing momenta, which are
very similar to those used in establishing Boltzmann's H-theorem.

Related to the above properties of the collision operator, we can introduce
a functional $\mathcal{Q}[g]$, such that Eq.~(\ref{trans2}) is equivalent to
finding its stationary point
\begin{equation}
\frac{\delta \mathcal{Q}[g]}{\delta g}=0.  \label{stat}
\end{equation}%
Specializing to the particle-hole symmetric case, cf.~ Eq.~(\ref{phsym}), the 
explicit form of the functional is
\begin{eqnarray}
&&\mathcal{Q}[g]=\frac{(2\pi )}{8v_{F}}\int \frac{d^{2}k}{(2\pi )^{2}}\frac{%
d^{2}k_{1}}{(2\pi )^{2}}\frac{d^{2}q}{(2\pi )^{2}}\Biggl\{  \notag \\
&&~~\frac{\delta (k-k_{1}-|\mathbf{k}+\mathbf{q}|+|\mathbf{k}_{1}-\mathbf{q}%
|)R_{1}(\mathbf{k},\mathbf{k}_{1},\mathbf{q})}{%
(e^{-v_{F}k/T}+1)(e^{v_{F}k_{1}/T}+1)(e^{v_{F}|\mathbf{k}+\mathbf{q}%
|/T}+1)(e^{-v_{F}|\mathbf{k}_{1}-\mathbf{q}|/T}+1)}  \notag \\
&&\times \left[ \frac{\mathbf{k}}{k}g(k,\omega )-\frac{\mathbf{k}_{1}}{k_{1}}%
g(k_{1},\omega )-\frac{(\mathbf{k}+\mathbf{q})}{|\mathbf{k}+\mathbf{q}|}g(|%
\mathbf{k}+\mathbf{q}|,\omega )+\frac{(\mathbf{k}_{1}-\mathbf{q})}{|\mathbf{k%
}_{1}-\mathbf{q}|}g(|\mathbf{k}_{1}-\mathbf{q}|,\omega )\right] ^{2}  \notag
\\
&&+\frac{\delta (k+k_{1}-|\mathbf{k}+\mathbf{q}|-|\mathbf{k}_{1}-\mathbf{q}%
|)R_{2}(\mathbf{k},\mathbf{k}_{1},\mathbf{q})}{%
(e^{-v_{F}k/T}+1)(e^{-v_{F}k_{1}/T}+1)(e^{v_{F}|\mathbf{k}+\mathbf{q}%
|/T}+1)(e^{v_{F}|\mathbf{k}_{1}-\mathbf{q}|/T}+1)}  \notag \\
&&\times \left[ \frac{\mathbf{k}}{k}g(k,\omega )+\frac{\mathbf{k}_{1}}{k_{1}}%
g(k_{1},\omega )-\frac{(\mathbf{k}+\mathbf{q})}{|\mathbf{k}+\mathbf{q}|}g(|%
\mathbf{k}+\mathbf{q}|,\omega )-\frac{(\mathbf{k}_{1}-\mathbf{q})}{|\mathbf{k%
}_{1}-\mathbf{q}|}g(|\mathbf{k}_{1}-\mathbf{q}|,\omega )\right] ^{2}\Biggr\}
\notag \\
&&~~~~~~~~~~~~~~~~~~~~~~~+\int \frac{d^{2}k}{(2\pi )^{2}}\frac{g(k,\omega
)[-i\omega g(k,\omega )/2-v_{F}/T]}{(e^{v_{F}k/T}+1)(e^{-v_{F}k/T}+1)}.
\label{func}
\end{eqnarray}

\subsection{Translational invariance and momentum conservation}

The translational invariance of the system immediately implies the presence of a 
zero mode of the operator ${\cal C}$, which corresponds to the shift of the 
distribution functions arising from changing to a linearly moving reference 
frame. The corresponding deviation $g_\lambda(k,\omega)$ has the form
\bea
\label{zeromode}
g_\lambda(k,\omega)=C(\omega)\psi_{tr}(k)\equiv C(\omega) k,
\eea
which is easily seen to annihilate the left hand side of (\ref{trans2}) due to 
momentum conservation. Note that this zero mode of the Boltzmann operator is 
orthogonal to any modes of the form (\ref{phsym}) which are the only ones that 
can be excited by an electric field at particle hole symmetry. This again 
expresses the fact that current and momentum are independent of each other at 
this special point. However, away from the Dirac point, or if a thermal gradient 
is applied instead of an electric field, the zero mode $\psi_{tr}$ will be 
excited by the driving field which leads to a diverging d.c. response in clean 
systems. This will be discussed in more detail in a forthcoming publication.

For the following we restrict to electrical conductivity at the particle-hole 
symmetric point where the above zero mode is irrelevant.

\subsection{Collinear limit}

\label{sec:coll}

In the previous analysis of a quantum Boltzmann equation for massless Dirac
fermions in two dimensions \cite{ssqhe}, it was noted that the phase space
for scattering of particles was logarithmically divergent in the collinear
limit. For the interaction considered in that paper, the collinear
scattering cross-section vanished, and so this singular phase space density
had no important consequences. The collinear scattering does not vanish for
the present Coulomb interaction, and so we need to consider this logarithmic
divergence seriously.

The physical origin of the divergent collinear scattering is related to the 
linear dispersion which implies that quasiparticles or -holes moving in the same 
direction share the same group velocity, independent of their energies. This 
leads to a diverging duration of collisions of nearly collinear particles, which 
is enhanced due to the low space dimensionality. To the extent that collinear 
scattering is very strong, and if we consider small enough frequencies, we may 
expect that quasiparticles and holes that move in the same direction in the 
plane will establish a pseudo-equilibrium characterized by an effective chemical 
potential and an effective temperature which will however depend on the 
direction of motion.

In linear response the deviations of these effective parameters from the 
equilibrium values $\mu$ and $T$ have to vary with ${\bf k}/k\cdot {\bf E}$ for 
symmetry reasons. Further, the effective temperature shift is easily shown to be 
identical to the mode $\psi_{tr}$ discussed above, and thus, it is thus ruled 
out at the particle hole symmetric point $\mu=0$. The remaining dominant mode of 
the function $g$ will correspond to an effective shift in chemical potential 
which translates into
\bea
\label{psimu}
g_\lambda(k,\omega)=  C(\omega) \psi_{\mu}(k)\equiv \frac{v_F}{T^2}\lambda 
C(\omega),
\eea
where the prefactor has been chosen so as to make $C(\omega)$ dimensionless.
With this Ansatz, which will be confirmed below, it simply remains to determine 
the prefactor $C(\omega)$, yielding the leading term in the non-equilibrium 
distribution. 
Note that the effective chemical potential shift ranges between $\pm C\,\hbar 
v_F e E/T$ depending on the direction. Comparing this to the temperature allows 
us to estimate the threshold electric field strength, $e E_{\rm lin}=T^2/\hbar 
v_F$, below which non-linear effects should remain small.

Let us now review in more detail, how the above physical picture arises in the 
formalism of the Boltzmann equation. The occurrence of a logarithmic divergence 
can be seen by allowing the incoming and outgoing momenta to be nearly 
collinear. Without loss of generality, we choose $%
\mathbf{k}=(k,0)$ with $k>0$. Also, we write $\mathbf{k_{1}}=(k_{1},k_{\perp
})$, $\mathbf{q}=(q,q_{\perp })$ with $k_{\perp }$ and $q_{\perp }$ small.
The divergence in the phase space density of the collision term proportional
to $R_{2}$ occurs when $k_{1}>0$, $k+q>0$ and $k_{1}-q>0$. 
Likewise, for scattering of oppositely charged particles a divergence occurs 
when their $k$ vectors are anticollinear which ensures collinear group 
velocities since ${\bf v}_{\bf k}=\lambda v_F {\bf k}/k$.
In this regime,
the argument of the energy conservation delta function of the particle-particle 
scattering term can be written as
\begin{eqnarray}
k+k_{1}-|\mathbf{k}+\mathbf{q}|-|\mathbf{k}_{1}-\mathbf{q}| &\approx &\frac{%
k_{\perp }^{2}}{2k_{1}}-\frac{q_{\perp }^{2}}{2(k+q)}-\frac{(k_{\perp
}-q_{\perp })^{2}}{2(k_{1}-q)}  \notag \\
&\equiv &-\frac{(k+k_{1})}{2(k+q)(k_{1}-q)}(q_{\perp }-\zeta _{1}k_{\perp
})(q_{\perp }-\zeta _{2}k_{\perp }),
\end{eqnarray}%
where $\zeta _{1,2}$ depend upon $k$, $k_{1}$, and $q$, and are the roots of
a quadratic equation which are defined by the expressions above. Then, the
phase space density for the $R_{2}$ term is proportional to
\begin{eqnarray}
&&\int dk_{\perp }dq_{\perp }\delta (k+k_{1}-|\mathbf{k}+\mathbf{q}|-|%
\mathbf{k}_{1}-\mathbf{q}|)=\int \frac{dk_{\perp }}{|k_{\perp }|}\frac{%
4(k+q)(k_{1}-q)}{(k+k_{1})|\zeta _{1}-\zeta _{2}|}  \notag \\
&&~~~~~~~~~~~~~~~~~=2\sqrt{\frac{k_{1}(k+q)(k_{1}-q)}{k}}\int \frac{%
dk_{\perp }}{|k_{\perp }|}\,.
\end{eqnarray}%
The logarithmic divergence as $k_{\perp }\rightarrow 0$ is now evident. This
divergence is clearly a consequence of the linear dispersion of the
fermions, and the above analysis also makes it clear that it is special to two
dimensions. As discussed in Ref.~\onlinecite{yaffe} for a similar divergence
in a different problem, we expect that this divergence is cutoff by
higher-order self energy corrections to the fermions. Such self-energy
corrections appear at order $\alpha $ in the perturbation theory, and so the
important range of the $k_{\perp }$ integral is between $T/v_{F}$ and $%
T\alpha /v_{F}$. So we may approximate \cite{yaffe}
\begin{equation}
\int \frac{dk_{\perp }}{|k_{\perp }|}\approx 2\ln (1/\alpha ),
\end{equation}%
and set $k_{\perp }=q_{\perp }=0$ elsewhere to obtain the leading
contribution to the collision integral in the limit $\alpha \rightarrow 0$.
Proceeding in this manner, the part of $\mathcal{C}$ on the right-hand side
of Eq.~(\ref{trans2}) proportional to $R_{2}$, which we denote $\mathcal{C}%
_{2}$, becomes
\begin{eqnarray}
\mathcal{C}_{2}[g] &\approx &-\frac{\ln (1/\alpha )}{2\pi ^{3}v_{F}}\frac{%
\mathbf{k}}{k}\int_{0}^{\infty }dk_{1}\int_{-k}^{k_{1}}\frac{dq}{q^{2}}\sqrt{%
\frac{k_{1}(k+q)(k_{1}-q)}{k}}  \notag \\
&~&~~~~~~\times \frac{R_{2}\left[ g(k,\omega )+g(k_{1},\omega )-g(k+q,\omega
)-g(k_{1}-q,\omega )\right] }{%
(e^{-v_{F}k/T}+1)(e^{-v_{F}k_{1}/T}+1)(e^{v_{F}(k+q)/T}+1)(e^{v_{F}(k_{1}-
q)/T}+1)%
}.  \label{trans3}
\end{eqnarray}%
Consonant with our discussion earlier in this subsection, 
a key property of the above expression for $\mathcal{C}_{2}$ was noted
by Kashuba~\cite{kashuba}: the function $g=\mbox{constant}$ is an
eigenvector of $\mathcal{C}_{2}$ with zero eigenvalue. The same is
also easily seen to apply to the portion ${\cal C}_1$ of $\mathcal{C}$ which is 
proportional to $R_{1}$.
Indeed, this is just the direction-specific chemical potential shift 
in Eq.~(\ref{psimu}), which naturally is a zero mode for collinear scattering, since it 
maintains a pseudo-equilibrium among particles moving in the same direction.

Going beyond the collinear limit, we conclude that there is an eigenvalue of 
$\mathcal{C}$
which is not proportional to $\ln (1/\alpha )$ in the limit of small $\alpha
$; the corresponding eigenvector is given by a constant $g(k)$ up to corrections 
of
order $[\ln (1/\alpha )]^{-1}$.

The solution of the Boltzmann equation in Eq.~(\ref{trans2}) requires that
we obtain the operator $\mathcal{C}^{-1}$, and the results above allow us to
constrain its form in the limit $\ln (1/\alpha) \gg 1$. Let $|\mu \rangle$
be the eigenvectors of $\mathcal{C}$ with eigenvalues $\lambda_\mu$. Then
\begin{equation}
\mathcal{C}^{-1} = \sum_{\mu} \frac{ |\mu \rangle \langle \mu |}{\lambda_\mu}\,,
\end{equation}
and in the limit of large $\ln (1/\alpha)$, $\mathcal{C}^{-1}$ is dominated \cite{kashuba}
by the eigenvector whose eigenvalue is not proportional to $\ln (1/\alpha)$.
Note that it is quite remarkable that in this limit we can solve the Boltzmann 
equation essentially exactly.

\subsection{Results}

From the reasoning in the previous subsection, we conclude that up to
corrections of order $[\ln (1/\alpha )]^{-1}$, we can choose $g$ to be of
the form
\begin{equation}
g(k,\omega )\approx \frac{v_{F}}{T^{2}}C(\omega ).  \label{gc}
\end{equation}%
We insert this parameterization into the functional $\mathcal{Q}[g]$ in Eq.~(%
\ref{func}); the solution of the stationarity condition in Eq.~(\ref{stat})
is then equivalent to requiring the vanishing of the derivative with
respect to $C$. We numerically evaluated the integrals in Eq.~(\ref{func})
using an elliptic co-ordinate system to solve the energy conservation
constraint \cite{ssqhe},  and obtained
\begin{equation*}
{\mathcal{Q}}[g]=\frac{1}{T}\frac{\ln {2}}{4\pi }\Biggl[\kappa \alpha
^{2}C^{2}(\omega )-2C(\omega )-i\left( \frac{\omega }{T}\right) C^{2}(\omega
)\Biggr]\,,
\end{equation*}%
with $\kappa =3.\,\allowbreak 646$ for the physical case $N=4$. From the
stationarity condition we then obtain
\begin{equation}
C(\omega )=\frac{1}{-i(\omega /T)+\kappa \alpha ^{2}}\,.
\end{equation}%
The conductivity can be obtained from $C(\omega )$ by combining Eqs.~(\ref%
{defj1}), (\ref{paramet}) and (\ref{gc}):
\begin{equation}
\sigma (\omega )=\frac{e^{2}}{h}\frac{Nk_{B}T\ln 2}{-i\hbar \omega +\kappa
k_{B}T\alpha ^{2}}\,,  \label{somega}
\end{equation}%
where we have re-inserted factors of $\hbar $ and $k_{B}$. Notice that the
conductivity depends only upon $\alpha (T)$, while all other factors of $%
v_{F}(T)$ cancel. Notice also the connection to the free particle result in
Eq.~(\ref{eqp1})---the only difference is that the infinitesimal $\eta $ has
been replaced by the inelastic relaxation energy $\kappa k_{B}T\alpha ^{2}$.

\section{Conclusions}

We conclude by briefly noting the conditions under which our main results
for conductivity in Eqs.~(\ref{somega}), (\ref{log}), and (\ref{power}) may
be observed in transport measurements. The key requirement is that $k_B T
\alpha^2$ be the largest infrared energy scale which quenches the ideal
Dirac fermion behavior. Thus, the sample size should be larger than the
inelastic scattering length $\ell_{\rm ee}\approx \hbar v_F/(k_B T \alpha^2)$. 
Similarly, the elastic
mean-free path from impurity scattering should be larger than $\ell_{\rm ee}$, 
too. Finally, particle-hole symmetry is also required, and so
the bias voltage should be smaller than $k_B T \alpha^2$.

It is possible to extend our analysis to include all the additional
perturbations noted in the previous paragraph, with a treatment of disorder effects
following that of Ref.~\onlinecite{aleiner}. When these perturbations are
weak (compared to $k_B T \alpha^2$), then in the collision-dominated regime,
a general hydrodynamic analysis is possible: this was presented recently in
Ref.~\onlinecite{mhd}. Also, in this regime the analysis of the Boltzmann 
equation greatly simplifies if the interactions are weak enough to ensure a 
strong logarithmic divergence in the collinear channel. The latter establishes 
pseudoequilibrium along different directions if the inelastic scattering time 
remains the shortest relevant time scale in the problem. Otherwise, a full 
analysis of the modified quantum
Boltzmann equation is required. These aspects will be discussed in future work.

\acknowledgments We thank S.~Das Sarma, F.~Guinea, P.~Kovtun, D.~E.~Sheehy, and
O.~Vafek for useful discussions. This research was supported by Deutsche
Forschungsgemeinschaft under grant FR 2627/1-1 (LF); by the Swiss National
Fund for Scientific Research under grant PA002-113151 (MM); by the NSF under
grant DMR-0537077 (SS), and by the Ames Laboratory, operated for the U.S.
Department of Energy by Iowa State University under Contract No.
DE-AC02-07CH11358 (JS).

\appendix

\section{Derivation of the Quantum Kinetic Equation}

An alternative derivation of the quantum kinetic equation can be carried out
in the framework of closed time contour ordered perturbation theory, as 
explicited in
Chapter 9 of Ref.~\cite{Kadanoff}. The problem we consider falls into the
generic class of a system describing particles interacting via a
distance-dependent density-density interaction. Thus the Hamiltonian is of
the form
\begin{eqnarray}\label{Hamiltonian}
H=H_{0}+\frac{1}{2}\int d^{2}\mathbf{r}d^{2}\mathbf{{r^{\prime }}}V(\mathbf{r%
},\mathbf{{r^{\prime }}})\rho (\mathbf{r})\rho (\mathbf{{r^{\prime }}})\;,
\end{eqnarray}%
where $\rho (\mathbf{r})$ denotes the particle density at spatial point $%
\mathbf{r}$. The following considerations are completely generic and apply
to any system which falls into the class of Hamiltonians presented in Eq.~%
\eqref{Hamiltonian}. The starting point of our discussion is given by
Eq.(9-7a) of Ref.~\cite{Kadanoff}, which has to be generalized to
incorporate a possible matrix structure of the Green's function (in our case
the Green's function lives in spinor space within a structure due to the N spin 
and valley species, and thus has a $2N\times 2N$
structure)
\begin{eqnarray}
\left[ \partial _{T}-\mathbf{\nabla }_{\mathbf{R}}U(\mathbf{R},T)\mathbf{%
\nabla }_{\mathbf{k}}\right] \mathbf{G}^{<}(\mathbf{k},\omega ;\mathbf{R}%
,T)=- &&\mathbf{G}^{<}(\mathbf{k},\omega ;\mathbf{R},T)\mathbf{\Sigma }^{>}(%
\mathbf{k},\omega ;\mathbf{R},T)  \notag \\
&+&\mathbf{G}^{>}(\mathbf{k},\omega ;\mathbf{R},T)\mathbf{\Sigma }^{<}(%
\mathbf{k},\omega ;\mathbf{R},T)
\end{eqnarray}%
where
\begin{eqnarray}
\mathbf{\Sigma }_{\alpha \beta }^{>,<}(\mathbf{r},t;\mathbf{R},T) &\approx
&-i^{2}\int d\mathbf{\overline{R}}\;d\mathbf{\overline{r}}\;V(\mathbf{R}+%
\mathbf{r}/2-\mathbf{\overline{R}}-\mathbf{\overline{r}}/2)\;V(\mathbf{R}-%
\mathbf{r}/2-\mathbf{\overline{R}}+\mathbf{\overline{r}}/2)\;\times   \notag
\label{kinetic} \\
&\times &\mathbf{G}_{\gamma \delta }^{<,>}(-\mathbf{\overline{r}},-t;R,T)%
\left[ \mathbf{G}_{\alpha \beta }^{>,<}(\mathbf{r},t;R,T)\mathbf{G}_{\delta
\gamma }^{>,<}(\mathbf{\overline{r}},t;R,T)\right.   \notag \\
&&-\left. \mathbf{G}_{\alpha \gamma }^{>,<}(\mathbf{R}+\mathbf{\overline{r}}%
/2-\mathbf{\overline{R}}+\mathbf{r}/2,t;R,T)\mathbf{G}_{\delta \beta }^{>,<}(%
\mathbf{R}+\mathbf{\overline{r}}/2-\mathbf{\overline{R}}+\mathbf{r}/2,t;R,T)%
\right]
\end{eqnarray}%
in the Born approximation (note: double indices are summed over). The
corresponding self-energy diagrams are the RPA-type contribution and the
maximally crossed diagram, see also Ref.~\cite{Kadanoff}. In a next step,
following the treatment in Kadanoff and Baym, we find that the Fourier
transform with respect to the relative coordinates (which corresponds to the
mixed Wigner transform) of Eq.~\eqref{kinetic} reads (note that in the
following we drop the dependence on the centre of mass coordinate $\mathbf{R}
$)
\begin{eqnarray}
\mathbf{\Sigma }_{\alpha \beta }^{>,<}({\mathbf{k}},\omega ;T) &=&\int \frac{%
d^{2}\mathbf{k_{1}}}{(2\pi )^{2}}\frac{d\omega _{1}}{2\pi }\frac{d^{2}%
\mathbf{{k_{2}}}}{(2\pi )^{2}}\frac{d\omega _{2}}{2\pi }\frac{d^{2}\mathbf{{%
k_{3}}}}{(2\pi )^{2}}\frac{d\omega _{3}}{2\pi }(2\pi )^{3}\delta (\mathbf{k}+%
\mathbf{k_{1}}-\mathbf{{k_{2}}}-\mathbf{{k_{3}}})\times   \notag
\label{eq:self} \\
&\times &\delta (\omega +\omega _{1}-\omega _{2}-\omega _{3})\times   \notag
\\
&\times &\left[ V(\mathbf{k}-\mathbf{{k_{2}}})V(\mathbf{k}-\mathbf{{k_{2}}})%
\mathbf{G}_{\gamma \delta }^{<,>}(\mathbf{k_{1}},\omega _{1})\mathbf{G}%
_{\alpha \beta }^{>,<}(\mathbf{{k_{2}}},\omega _{2})\mathbf{G}_{\delta
\gamma }^{>,<}(\mathbf{{k_{3}}},\omega _{3})\right.   \notag \\
&-&\left. V(\mathbf{k}-\mathbf{{k_{2}}})V(\mathbf{k}-\mathbf{{k_{3}}})%
\mathbf{G}_{\gamma \delta }^{<,>}(\mathbf{k},\omega _{1})\mathbf{G}_{\alpha
\gamma }^{>,<}(\mathbf{{k_{2}}},\omega _{2})\mathbf{G}_{\delta \beta }^{>,<}(%
\mathbf{{k_{3}}},\omega _{3})\right] \;.
\end{eqnarray}%
Until now all the formulae are completely generic and not specific to
graphene. In order to make connection to the problem of graphene we note
that the Green's function of the spinors $\Psi $ is related to the Green's
function of the $\gamma $ through
\begin{equation*}
\mathbf{G}^{<,>}(\mathbf{k},\omega )=U_{\mathbf{k}}^{-1}g^{<,>}(\mathbf{k}%
,\omega )U_{\mathbf{k}}
\end{equation*}%
where the unitary matrix $U_{\mathbf{k}}^{-1}$ according to Eq.~%
\eqref{eq:unitary} is given by
\begin{equation*}
U_{\mathbf{k}}^{-1}=\frac{1}{\sqrt{2}k}\left(
\begin{array}{cc}
k & k \\
K & -K%
\end{array}%
\right) \;.
\end{equation*}%
Furthermore we note, that the summation over spin and valley indices only
affects the RPA-like diagram, which thus receives a prefactor $N$ and the
resulting matrix equation is an equation, whose indices only carry over the $%
2\times 2$ spinor. This allows to rewrite Eq.~\eqref{eq:self} as
\begin{eqnarray}
\mathbf{\Sigma }_{\alpha \beta }^{>,<}({\mathbf{k}},\omega ;T) &=&\int \frac{%
d^{2}\mathbf{k_{1}}}{(2\pi )^{2}}\frac{d\omega _{1}}{2\pi }\frac{d^{2}%
\mathbf{{k_{2}}}}{(2\pi )^{2}}\frac{d\omega _{2}}{2\pi }\frac{d^{2}\mathbf{{%
k_{3}}}}{(2\pi )^{2}}\frac{d\omega _{3}}{2\pi }(2\pi )^{3}\delta (\mathbf{k}+%
\mathbf{k_{1}}-\mathbf{{k_{2}}}-\mathbf{{k_{3}}})\times   \notag \\
&\times &\delta (\omega +\omega _{1}-\omega _{2}-\omega _{3})\times   \notag
\\
&\times &\left[ NV(\mathbf{k}-\mathbf{{k_{2}}})V(\mathbf{k}-\mathbf{{k_{2}}}%
)\right. \times   \notag \\
&\times &\left. \left( U_{\mathbf{k_{1}}}^{-1}g^{<,>}(\mathbf{k_{1}},\omega
_{1})U_{\mathbf{k_{1}}}\right) _{\gamma \delta }\left( U_{\mathbf{{k_{2}}}%
}^{-1}g^{>,<}(\mathbf{{k_{2}}},\omega _{2})U_{\mathbf{{k_{2}}}}\right)
_{\alpha \beta }\left( U_{\mathbf{{k_{3}}}}^{-1}g^{>,<}(\mathbf{{k_{3}}}%
,\omega _{3})U_{\mathbf{{k_{3}}}}\right) _{\delta \gamma }\right.   \notag \\
&-&\left. V(\mathbf{k}-\mathbf{{k_{2}}})V(\mathbf{k}-\mathbf{{k_{3}}}%
)\right.   \notag \\
&\times &\left. \left( U_{\mathbf{k_{1}}}^{-1}g^{<,>}(\mathbf{k_{1}},\omega
_{1})U_{\mathbf{k_{1}}}\right) _{\gamma \delta }\left( U_{\mathbf{{k_{2}}}%
}^{-1}g^{>,<}(\mathbf{{k_{2}}},\omega _{2})U_{\mathbf{{k_{2}}}}\right)
_{\alpha \gamma }\left( U_{\mathbf{{k_{3}}}}^{-1}g^{>,<}(\mathbf{{k_{3}}}%
,\omega _{3})U_{\mathbf{{k_{3}}}}\right) _{\delta \beta }\right]
\end{eqnarray}%
Accounting for the fact that the operators $\gamma $ describe sharp
quasiparticles the lesser and greater Green's functions are given by
\begin{equation*}
g_{\lambda \lambda ^{\prime }}^{<}(\mathbf{k},\omega ;T)=2\pi \delta (\omega
-\epsilon _{\lambda }(\mathbf{k},T))f_{\lambda }(\mathbf{k},T)\delta
_{\lambda ,\lambda ^{\prime }}
\end{equation*}%
and
\begin{equation*}
g_{\lambda \lambda ^{\prime }}^{>}(\mathbf{k},\omega ;T)=2\pi \delta (\omega
-\epsilon _{\lambda }(\mathbf{k},T))\left[ 1-f_{\lambda }(\mathbf{k},T)%
\right] \delta _{\lambda ,\lambda ^{\prime }}\;,
\end{equation*}%
where we assumed the distribution function of the quasiparticles to have no
off-diagonal components, which is justified to linear order in the potential
gradient. We can formulate the kinetic equation for the diagonal part of the
distribution function as
\begin{eqnarray}
\left[ \partial _{T}-\nabla _{\mathbf{R}}U(\mathbf{R},T)\nabla _{\mathbf{k}}%
\right] f_{\mu }(\mathbf{k},T)=- &&f_{\mu }(\mathbf{k},T)\left( U_{\mathbf{k}%
}\mathbf{\Sigma }^{>}({\mathbf{k}},\omega =\epsilon _{\mu }(\mathbf{k});T)U_{%
\mathbf{k}}^{-1}\right) _{\mu \mu }  \notag \\
&+&\left[ 1-f_{\mu }(\mathbf{k},T)\right] \left( U_{\mathbf{k}}\mathbf{%
\Sigma }^{<}({\mathbf{k}},\omega =\epsilon _{\mu }(\mathbf{k});T)U_{\mathbf{k%
}}^{-1}\right) _{\mu \mu }
\end{eqnarray}%
or equivalently
\begin{eqnarray}
\left[ \partial _{T}-\nabla _{\mathbf{R}}U(\mathbf{R},T)\nabla _{\mathbf{k}}%
\right] f_{\mu }(\mathbf{k},T)=- &&f_{\mu }(\mathbf{k},T)\mathbf{\sigma }%
_{\mu \mu }^{>}({\mathbf{k}},\omega =\epsilon _{\mu }(\mathbf{k});T)  \notag
\label{transeq} \\
&+&\left[ 1-f_{\mu }(\mathbf{k},T)\right] \mathbf{\sigma }_{\mu \mu }^{<}({%
\mathbf{k}},\omega =\epsilon _{\mu }(\mathbf{k});T)\;.
\end{eqnarray}%
Exploiting the form of the lesser and greater Green's functions we can
rewrite the self-energies as (note that $\mu $, in contrast to the other
double indices, is not summed over here and subsequently)
\begin{eqnarray}
&&\mathbf{\sigma }_{\mu \mu }^{>}({\mathbf{k}},\omega =\epsilon _{\mu }(%
\mathbf{k});T)=\int \frac{d^{2}\mathbf{k_{1}}}{(2\pi )^{2}}\frac{d\omega _{1}%
}{2\pi }\frac{d^{2}\mathbf{{k_{2}}}}{(2\pi )^{2}}\frac{d\omega _{2}}{2\pi }%
\frac{d^{2}\mathbf{{k_{3}}}}{(2\pi )^{2}}\frac{d\omega _{3}}{2\pi }(2\pi
)^{3}\delta (\mathbf{k}+\mathbf{k_{1}}-\mathbf{{k_{2}}}-\mathbf{{k_{3}}})
\notag \\
&\times &\delta (\epsilon _{\mu }(\mathbf{k})+\omega _{1}-\omega _{2}-\omega
_{3})\times   \notag \\
&\times &\left[ NV(\mathbf{k}-\mathbf{{k_{2}}})V(\mathbf{k}-\mathbf{{k_{2}}}%
)(2\pi )^{3}\delta (\omega _{1}-\epsilon _{\lambda }(\mathbf{k_{1}}))\delta
(\omega _{2}-\epsilon _{\lambda _{1}}(\mathbf{{k_{2}}}))\delta (\omega
_{3}-\epsilon _{\lambda _{2}}(\mathbf{{k_{3}}}))\right.   \notag \\
&\times &\left. M_{\lambda _{2}\lambda }(\mathbf{{k_{3}}},\mathbf{k_{1}}%
)M_{\lambda \lambda _{2}}(\mathbf{k_{1}},\mathbf{{k_{3}}})M_{\mu \lambda
_{1}}(\mathbf{k},\mathbf{{k_{2}}})M_{\lambda _{1}\mu }(\mathbf{{k_{2}}},%
\mathbf{k})f_{\lambda }(\mathbf{k_{1}},T)(1-f_{\lambda _{1}}(\mathbf{{k_{2}}}%
,T))(1-f_{\lambda _{2}}(\mathbf{{k_{3}}},T))\right.   \notag \\
&-&\left. V(\mathbf{k}-\mathbf{{k_{2}}})V(\mathbf{k}-\mathbf{{k_{3}}})(2\pi
)^{3}\delta (\omega _{1}-\epsilon _{\lambda }(\mathbf{k_{1}}))\delta (\omega
_{2}-\epsilon _{\lambda _{1}}(\mathbf{{k_{2}}}))\delta (\omega _{3}-\epsilon
_{\lambda _{2}}(\mathbf{{k_{3}}}))\right.   \notag \\
&\times &\left. M_{\lambda \lambda _{2}}(\mathbf{k_{1}},\mathbf{{k_{3}}}%
)M_{\lambda _{1}\lambda }(\mathbf{{k_{2}}},\mathbf{k_{1}})M_{\mu \lambda
_{1}}(\mathbf{k},\mathbf{{k_{2}}})T_{\lambda _{2}\mu }(\mathbf{{k_{3}}},%
\mathbf{k})f_{\lambda }(\mathbf{k_{1}},T)(1-f_{\lambda _{1}}(\mathbf{{k_{2}}}%
,T))(1-f_{\lambda _{2}}(\mathbf{{k_{3}}},T))\right] \nonumber \\
\end{eqnarray}%
and
\begin{eqnarray}
&&\mathbf{\sigma }_{\mu \mu }^{<}({\mathbf{k}},\omega =\epsilon _{\mu }(%
\mathbf{k});\mathbf{R},T)=\int \frac{d^{2}\mathbf{k_{1}}}{(2\pi )^{2}}\frac{%
d\omega _{1}}{2\pi }\frac{d^{2}\mathbf{{k_{2}}}}{(2\pi )^{2}}\frac{d\omega
_{2}}{2\pi }\frac{d^{2}\mathbf{{k_{3}}}}{(2\pi )^{2}}\frac{d\omega _{3}}{%
2\pi }(2\pi )^{3}\delta (\mathbf{k}+\mathbf{k_{1}}-\mathbf{{k_{2}}}-\mathbf{{%
k_{3}}})  \notag \\
&\times &\delta (\epsilon _{\mu }(\mathbf{k})+\omega _{1}-\omega _{2}-\omega
_{3})\times   \notag \\
&\times &\left[ NV(\mathbf{k}-\mathbf{{k_{2}}})V(\mathbf{k_{1}}-\mathbf{{%
k_{2}}})(2\pi )^{3}\delta (\omega _{1}-\epsilon _{\lambda }(\mathbf{k_{1}}%
))\delta (\omega _{2}-\epsilon _{\lambda _{1}}(\mathbf{{k_{2}}}))\delta
(\omega _{3}-\epsilon _{\lambda _{2}}(\mathbf{{k_{3}}}))\right.   \notag \\
&\times &\left. M_{\lambda _{2}\lambda }(\mathbf{{k_{3}}},\mathbf{k_{1}}%
)M_{\lambda \lambda _{2}}(\mathbf{k_{1}},\mathbf{{k_{3}}})M_{\mu \lambda
_{1}}(\mathbf{k},\mathbf{{k_{2}}})M_{\lambda _{1}\mu }(\mathbf{{k_{2}}},%
\mathbf{k})(1-f_{\lambda }(\mathbf{k_{1}},T))f_{\lambda _{1}}(\mathbf{{k_{2}}%
},T)f_{\lambda _{2}}(\mathbf{{k_{3}}},T)\right.   \notag \\
&-&\left. V(\mathbf{k}-\mathbf{{k_{2}}})V(\mathbf{k}-\mathbf{{k_{3}}})(2\pi
)^{3}\delta (\omega _{1}-\epsilon _{\lambda }(\mathbf{k_{1}}))\delta (\omega
_{2}-\epsilon _{\lambda _{1}}(\mathbf{{k_{2}}}))\delta (\omega _{3}-\epsilon
_{\lambda _{2}}(\mathbf{{k_{3}}}))\right.   \notag \\
&\times &\left. M_{\lambda \lambda _{2}}(\mathbf{k_{1}},\mathbf{{k_{3}}}%
)M_{\lambda _{1}\lambda }(\mathbf{{k_{2}}},\mathbf{k_{1}})M_{\mu \lambda
_{1}}(\mathbf{k},\mathbf{{k_{2}}})M_{\lambda _{2}\mu }(\mathbf{{k_{3}}},%
\mathbf{k})(1-f_{\lambda }(\mathbf{k_{1}},T))f_{\lambda _{1}}(\mathbf{{k_{2}}%
},T)f_{\lambda _{2}}(\mathbf{{k_{3}}},T)\right] \;, \nonumber \\
\end{eqnarray}%
where the shorthand notation
\begin{equation*}
M_{\lambda \lambda _{1}}(\mathbf{k},\mathbf{k_{1}})=\left[ U_{\mathbf{k}}^{%
\phantom{1}}U_{\mathbf{k_{1}}}^{-1}\right] _{\lambda \lambda _{1}}=\frac{1}{2%
}\left(
\begin{array}{cc}
1+\frac{K^{\star }K_{1}}{kk_{1}} & 1-\frac{K^{\star }K_{1}}{kk_{1}} \\
1-\frac{K^{\star }K_{1}}{kk_{1}} & 1+\frac{K^{\star }K_{1}}{kk_{1}}%
\end{array}%
\right) _{\lambda \lambda _{1}}=\frac{1}{2}\left( 1+\lambda \lambda _{1}%
\frac{K^{\star }K_{1}}{kk_{1}}\right)
\end{equation*}%
was introduced. The connection with the matrix elements defined in 
Eq.~\eqref{deft}
can be easily established and reads
\begin{eqnarray}\label{melement}
T_{\lambda \lambda_1 \lambda_2 \lambda_3} (\vk, \vko,{\bf q})=\frac{1}{2} V(-
{\bf q}) M_{\lambda \lambda_3}(\vk +{\bf q},\vk) M_{\lambda_1 \lambda_2}(\vko-
{\bf q},\vko)\;.
\end{eqnarray}
Using Eq.~\eqref{transeq} and preforming a sequence of
transformations finally yields
\begin{eqnarray}
&&\left[ \partial _{T}-\nabla _{\mathbf{R}}U(\mathbf{R},T)\nabla _{\mathbf{k}%
}\right] f_{\mu }(\mathbf{k},T)=\frac{2\pi }{v_{F}}\int \frac{d^{2}\mathbf{%
k_{1}}}{(2\pi )^{2}}\frac{d^{2}\mathbf{q}}{(2\pi )^{2}}\delta (\mu k+\lambda
k_{1}-\lambda _{1}|\mathbf{k}+\mathbf{q}|-\lambda _{2}|\mathbf{k_{1}}-%
\mathbf{q}|)  \notag \\
&\times &\left[ NV(-\mathbf{q})V(-\mathbf{q})M_{\lambda _{2}\lambda }(%
\mathbf{k_{1}}-\mathbf{q},\mathbf{k_{1}})M_{\lambda \lambda _{2}}(\mathbf{%
k_{1}},\mathbf{k_{1}}-\mathbf{q})M_{\mu \lambda _{1}}(\mathbf{k},\mathbf{q}+%
\mathbf{k})M_{\lambda _{1}\mu }(\mathbf{q}+\mathbf{k},\mathbf{k})\right.
\notag \\
&-&\left. V(-\mathbf{q})V(\mathbf{k}-\mathbf{k_{1}}+\mathbf{q})M_{\lambda
\lambda _{2}}(\mathbf{k_{1}},\mathbf{k_{1}}-\mathbf{q})M_{\lambda
_{1}\lambda }(\mathbf{q}+\mathbf{k},\mathbf{k_{1}})M_{\mu \lambda _{1}}(%
\mathbf{k},\mathbf{q}+\mathbf{k})M_{\lambda _{2}\mu }(\mathbf{k_{1}}-\mathbf{%
q},\mathbf{k})\right]   \notag \\
&&\left[ (1-f_{\mu }(\mathbf{k},T))(1-f_{\lambda }(\mathbf{k_{1}}%
,T))f_{\lambda _{1}}(\mathbf{q}+\mathbf{k},T)f_{\lambda _{2}}(\mathbf{k_{1}}-%
\mathbf{q},T)\right.   \notag \\
&-&\left. f_{\mu }(\mathbf{k},T)f_{\lambda }(\mathbf{k_{1}},T)(1-f_{\lambda
_{1}}(\mathbf{q}+\mathbf{k},T))(1-f_{\lambda _{2}}(\mathbf{k_{1}}-\mathbf{q}%
,T))\right] \;.
\end{eqnarray}%
In a next step we will make connection to the Golden rule result of the main
text. It is straightforward to see that using Eq.~\eqref{melement} we can
rewrite the above expression to yield
\begin{eqnarray}
&&\left[ \partial _{T}-\nabla _{\mathbf{R}}U(\mathbf{R},T)\nabla _{\mathbf{k}%
}\right] f_{\mu }(\mathbf{k},T)=\frac{2\pi }{v_{F}}\int \frac{d^{2}\mathbf{%
k_{1}}}{(2\pi )^{2}}\frac{d^{2}\mathbf{q}}{(2\pi )^{2}}\delta (\mu k+\lambda
k_{1}-\lambda _{1}|\mathbf{k}+\mathbf{q}|-\lambda _{2}|\mathbf{k_{1}}-%
\mathbf{q}|)  \notag \\
&\times &4\left[ N|T_{\mu \lambda \lambda _{2}\lambda _{1}}(\mathbf{k},%
\mathbf{k_{1}},\mathbf{q})|^{2}-T_{\mu \lambda \lambda _{2}\lambda _{1}}(%
\mathbf{k},\mathbf{k_{1}},\mathbf{q})T_{\mu \lambda \lambda _{1}\lambda
_{2}}^{\star }(\mathbf{k},\mathbf{k_{1}},\mathbf{k_{1}}-\mathbf{k}-\mathbf{q}%
)\right]   \notag \\
&&\left[ (1-f_{\mu }(\mathbf{k},T))(1-f_{\lambda }(\mathbf{k_{1}}%
,T))f_{\lambda _{1}}(\mathbf{q}+\mathbf{k},T)f_{\lambda _{2}}(\mathbf{k_{1}}-%
\mathbf{q},T)\right.   \notag \\
&-&\left. f_{\mu }(\mathbf{k},T)f_{\lambda }(\mathbf{k_{1}},T)(1-f_{\lambda
_{1}}(\mathbf{q}+\mathbf{k},T))(1-f_{\lambda _{2}}(\mathbf{k_{1}}-\mathbf{q}%
,T))\right]
\end{eqnarray}%
Energy and momentum conservation restricts the valid combinations of
particles and holes scattering, see Ref.~\cite{ssbook,ssqhe}, which
simplifies the above expression. Applying all these simplifications and
shifting the variables appropriately we obtain
\begin{eqnarray}
&&\left[ \partial _{T}-\nabla _{\mathbf{R}}U(\mathbf{R},T)\nabla _{\mathbf{k}%
}\right] f_{\mu }(\mathbf{k},T)=-\frac{(2\pi )}{v_{F}}\int \frac{d^{2}k_{1}}{%
(2\pi )^{2}}\frac{d^{2}k_{2}}{(2\pi )^{2}}\Biggl\{  \notag \\
&&\delta (k-k_{1}-|\mathbf{k}+\mathbf{q}|+|\mathbf{k}_{1}-\mathbf{q}|)%
\overline{R}_{1}\Bigl\{f_{\mu }(\mathbf{k},t)f_{-\mu }(\mathbf{k}%
_{1},t)[1-f_{\mu }(\mathbf{k}+\mathbf{q},t)][1-f_{-\mu }(\mathbf{k}_{1}-%
\mathbf{q},t)]  \notag \\
&&~~~~~~~~~~~~~~~~~~~~~~~~~~~~~~~-[1-f_{\mu }(\mathbf{k},t)][1-f_{-\mu }(%
\mathbf{k}_{1},t)]f_{\mu }(\mathbf{k}+\mathbf{q},t)f_{-\mu }(\mathbf{k}_{1}-%
\mathbf{q},t)\Bigr\}  \notag \\
&&\delta (k+k_{1}-|\mathbf{k}+\mathbf{q}|-|\mathbf{k}_{1}-\mathbf{q}|)%
\overline{R}_{2}\Bigl\{f_{\mu }(\mathbf{k},t)f_{\mu }(\mathbf{k}%
_{1},t)[1-f_{\mu }(\mathbf{k}+\mathbf{q},t)][1-f_{\mu }(\mathbf{k}_{1}-%
\mathbf{q},t)]  \notag \\
&&~~~~~~~~~~~~~~~~~~~~~~~~~~~~~~~-[1-f_{\mu }(\mathbf{k},t)][1-f_{\mu }(%
\mathbf{k}_{1},t)]f_{\mu }(\mathbf{k}+\mathbf{q},t)f_{\mu }(\mathbf{k}_{1}-%
\mathbf{q},t)\Bigr\}\Biggr\},
\end{eqnarray}%
where
\begin{eqnarray}
\overline{R}_{1} &=&4N\left( |T_{+--+}(\mathbf{k},\mathbf{k_{1}},\mathbf{q}%
)|^{2}+|T_{+-+-}(\mathbf{k},\mathbf{k_{1}},\mathbf{k_{1}}-\mathbf{k}-\mathbf{%
q})|^{2}\right)   \notag \\
&&-4T_{+--+}(\mathbf{k},\mathbf{k_{1}},\mathbf{q})T_{+-+-}^{\star }(\mathbf{k%
},\mathbf{k_{1}},\mathbf{k_{1}}-\mathbf{k}-\mathbf{q})  \notag \\
&&-4T_{+-+-}^\star(\mathbf{k},\mathbf{k_{1}},\mathbf{k_{1}}-\mathbf{k}-
\mathbf{q}%
)T_{+--+}(\mathbf{k},\mathbf{k_{1}},\mathbf{q})
\end{eqnarray}%
and
\begin{eqnarray}
\overline{R}_{2} &=&4N|T_{++++}(\mathbf{k},\mathbf{k_{1}},\mathbf{q})|^{2}
\notag \\
&&-4T_{++++}(\mathbf{k},\mathbf{k_{1}},\mathbf{q})T_{++++}^{\star }(\mathbf{k%
},\mathbf{k_{1}},\mathbf{k_{1}}-\mathbf{k}-\mathbf{q})\;.
\end{eqnarray}%
Performing the appropriate shifts allows to write
\begin{eqnarray}
\overline{R}_{1} &=&4(N-1)|T_{+--+}(\mathbf{k},\mathbf{k_{1}},\mathbf{q}%
)|^{2}+4(N-1)|T_{+-+-}(\mathbf{k},\mathbf{k_{1}},\mathbf{k_{1}}-\mathbf{k}-%
\mathbf{q})|^{2}  \notag \\
&&+4|T_{+--+}(\mathbf{k},\mathbf{k_{1}},\mathbf{q})-T_{+-+-}(\mathbf{k},%
\mathbf{k_{1}},\mathbf{k_{1}}-\mathbf{k}-\mathbf{q})|^{2}
\end{eqnarray}%
and
\begin{eqnarray}
\overline{R}_{2} &=&4(N-1)|T_{++++}(\mathbf{k},\mathbf{k_{1}},\mathbf{q}%
)|^{2}  \notag \\
&&+2|T_{++++}(\mathbf{k},\mathbf{k_{1}},\mathbf{q})-T_{++++}(\mathbf{k},%
\mathbf{k_{1}},\mathbf{k_{1}}-\mathbf{k}-\mathbf{q})|^{2}\;,
\end{eqnarray}%
which establishes the equivalence of Fermi's Golden rule and the Keldysh
treatment, see Eq.~\eqref{deft12}.

\end{document}